\documentstyle[12pt]{article}

\begin{document}
\title{Time dependent transformations in deformation quantization}
\author{
Nuno Costa Dias\footnote{{\it ncdias@mail.telepac.pt}} \\
Jo\~ao Nuno Prata\footnote{{\it joao.prata@ulusofona.pt}} \\
{{\it Departamento de Matem\'{a}tica}} \\
{{\it Universidade Lus\'ofona de Humanidades e Tecnologias}} \\
{{\it Av. Campo Grande, 376, 1749-024 Lisboa, Portugal}}}
\date{}
\maketitle

\begin{abstract}
We study the action of time dependent canonical and coordinate transformations in phase space quantum mechanics. We extend the covariant formulation of the theory by providing a formalism that is fully invariant under both standard and time dependent coordinate transformations. This result considerably enlarges the set of possible phase space representations of quantum mechanics and makes it possible to construct a causal representation for the distributional sector of Wigner quantum mechanics.
\end{abstract}

\section{Introduction}

The phase space formulation of quantum mechanics was originally introduced by Weyl \cite{Weyl} and Wigner \cite{Wigner} and further developed by Moyal \cite{Moyal}. The theory lives on the classical phase space and its key algebraic structures (the star-product and the Moyal bracket) are both $\hbar$-deformations of the standard algebraic structures of classical mechanics \cite{Wigner2}-\cite{Flato1}. Because of this its mathematical formalism is remarkably similar to that of classical statistical mechanics, a property that has been perceived by many as a conceptual and technical advantage when addressing a wide range of specific problems \cite{Flato1}-\cite{nuno2}. This relative success, together with the fact that the deformed algebraic structures play a key part in some current developments in M-theory \cite{Fairlie2}-\cite{Pinzul} led to an intense research on applications of the deformation quantization approach as well as on the further development of its mathematical structure.

The Wigner theory uses the symmetric ordering prescription (the Weyl order) to find a particular phase space representation of quantum mechanics. Different representations provide different points of view and may suggest new solutions for both technical and conceptual problems. They may even suggest new interpretations for the entire quantum theory, as in the case of the De Broglie Bohm formulation. 
For its importance, the topic of finding new, more general phase space representations of quantum mechanics has been studied in depth. Cohen \cite{Cohen} introduced a generalization of the Weyl map providing in an unified fashion all phase space representations that correspond to different ordering prescriptions of operator quantum mechanics. The resulting theory of quasi-distributions includes as particular cases the Wigner and the De Broglie Bohm \cite{nuno1,nuno9} formulations. Vey \cite{Vey} and several others \cite{nuno6,Flato1}, \cite{Wilde}-\cite{Fedosov2} developed the covariant generalization of Wigner's theory. The new formulation renders phase space quantum mechanics fully invariant under the action of phase space coordinate transformations. By doing so it provides a general formula for the $\hbar$-deformations of the Poisson bracket and makes it possible to apply deformation quantization methods to a larger set of dynamical systems including some of those displaying the structure of a curved phase space manifold. 

The aim of this paper is to extend the covariant formulation further by admitting the possibility of time dependent coordinate transformations. We will study the action of these transformations in phase space quantum mechanics and re-write the covariant Wigner theory in a fully invariant form under their action. We derive the time dependent covariant form of the starproduct, Moyal bracket, Moyal dynamical equation and stargenvalue equation as well as the covariant probabilistic functionals, providing all key ingredients of the time dependent covariant formulation of Wigner quantum mechanics.
This result enlarges the set of possible phase space representations of quantum mechanics and provides a more general formula for the $\hbar$-deformations of the Poisson bracket, which may now include an explicit time dependence.

The new set of representations provides new possible formulations for a generic quantum mechanical problem. In some cases this may considerably simplify the technical resolution of the problem. In section 6 a simple example illustrates how a suitable time dependent representation leads to a far simpler description of the dynamics of the quasidistribution.
More relevant is the fact that the new formalism makes it possible to construct a phase space representation of quantum mechanics where the quasidistribution displays a classical causal structure, i.e. the Wigner function evolves according to the Liouville equation. 
A formulation displaying this set of properties cannot be (easily) accomplished using the standard methods of quantum mechanics (not even within the standard covariant Wigner formalism) and it proves that the De Broglie Bohm theory is not the unique possible causal formulation of quantum mechanics. In the new causal representation the quantum dynamical behavior is completely removed from the distributional sector of the theory and is exclusively placed on the observables' sector. In particular if the quasidistribution is positive defined at the initial time it will remain so for all times. These properties reinforce the formal analogy between phase space quantum mechanics and classical statistical mechanics and make the causal formulation especially suitable to study the semiclassical limit of quantum mechanics.
  
This paper is organized as follows: in section 2 we review the main topics of the covariant formulation of the Wigner theory. In section 3 we study the action of time dependent canonical transformations in standard operator quantum mechanics. Particular attention is devoted to the behavior of the density matrix. In section 4 we derive the time dependent covariant formulation of Wigner quantum mechanics. In section 5 a particular set of coordinates is used to obtain the causal phase space representation. In section 6 a simple example illustrates some of the former results and in section 7 we present the conclusions.

\section{Covariant Wigner quantum mechanics}

Let us consider a $N$ dimensional dynamical system. Its classical formulation lives on the phase space $T^*M$ which, to make it simple, we assume to be flat. A global Darboux chart can then be naturally defined on $T^*M$, for which the symplectic structure reads $w=d q_i \wedge d p_i$, where $\{q_i,p_i, i=1..N\}$ is a set of canonical variables.

Upon quantization the set $\{\hat q_i\}$ yields a complete set of commuting observables. Let then $\hat A(\hat{\vec q}, \hat{\vec p})$ be a generic operator acting on the physical Hilbert space ${\cal H}$. The Weyl map:
\begin{equation}
W_{(q,p)}(\hat A)= \hbar^N \int d^N\vec y e^{-i\vec p \cdot \vec y} <\vec q+\frac{\hbar}{2} \vec y|\hat A|\vec q-\frac{\hbar}{2} \vec y>,
\end{equation}
where we introduced the vector notation $\vec y \equiv (y_1, \cdots, y_N)$ and $|\vec q \pm \frac{\hbar}{2} \vec y>$ are eigenstates of $\hat{\vec q}$;
provides a Lie algebra isomorphism between the algebra $\hat{\cal A}({\cal H})$ of linear operators acting on the Hilbert space ${\cal H}$ and the algebra of phase space functions ${\cal A}(T^*M)$ endowed with a $*$-product and Moyal bracket (let $A, B \in {\cal A}(T^*M)$): 
\begin{equation}
A*_{(q,p)} B = A e^{\frac{i \hbar}{2} {\buildrel { \leftarrow}\over\partial}_k J_{(q,p)}^{kl} {\buildrel { \rightarrow}\over\partial}_l} B, \qquad \left[ A, B \right]_{M_{(q,p)}} = \frac{2}{\hbar}  A \sin \left(\frac{\hbar}{2}  {\buildrel { \leftarrow}\over\partial}_k J_{(q,p)}^{kl} {\buildrel { \rightarrow}\over\partial}_l \right) B,
\end{equation}
where the derivatives ${\buildrel { \leftarrow}\over\partial}$ and ${\buildrel { \rightarrow}\over\partial}$ act on $A$ and $B$, respectively and $J_{(q,p)}^{kl}$ is the $kl$-th element of the symplectic matrix in the variables $(\vec q,\vec p)$:
\begin{equation}
J_{(q,p)} = \left(
\begin{array}{l r}
0_{N \times N} & - 1_{N \times N}\\
1_{N \times N} & 0_{N \times N}
\end{array}
\right).
\end{equation}
We also introduced the compact notation: $O^k = p_k,  k=1, \cdots ,N$; $O^k = q_{k-N},  k=N+1, \cdots, 2N$; ${\partial} / {\partial O^k} = \partial_k$ and sum over repeated indices is understood.

We now consider a second set of fundamental operators  $(\hat{\vec Q}, \hat{\vec P})$ unitarily related to
$(\hat{\vec q},\hat{\vec p})$, i.e. $\hat{\vec q}=\hat U \hat{\vec Q}\hat U^{-1}$ and $\hat{\vec p}=\hat U \hat{\vec P}\hat U^{-1}$ where $\hat U$ is some unitary operator. The new operators satisfy the Heisenberg commutation relations, yield a new Weyl map $W_{(Q,P)}$ and induce a phase space transformation $(\vec q,\vec p) \longrightarrow (\vec Q, \vec P)$ acting on a generic observable through the procedure (let $U= W_{(Q,P)} (\hat U)$):
\begin{equation}
A(\vec q,\vec p)=W_{(q,p)} (\hat A) \longrightarrow A'(\vec Q,\vec P)=W_{(Q,P)} (\hat A) = U *_{(Q,P)} A(\vec Q,\vec P) *_{(Q,P)} U^{-1}\; .
\end{equation}
The phase space implementation of the unitary transformation preserves the starproduct and the Moyal bracket but, as is well known, it does not act as a coordinate transformation (the exceptions are the linear transformations): let
$\vec q(\vec Q,\vec P)=W_{(Q,P)}(\hat q)$ and $\vec p(\vec Q,\vec P)=W_{(Q,P)} (\hat p)$ and we find that in general $A'(\vec Q,\vec P) \not= A(\vec q(\vec Q,\vec P),\vec p(\vec Q,\vec P))$. We conclude that the standard Wigner formulation is non-covariant.

We now introduce the generalized Weyl map \cite{nuno6}. Let the transformation $(\vec q,\vec p) \longrightarrow (\vec Q, \vec P)$ be a phase space diffeomorphism defined, in general terms, by $\vec q=\vec q(\vec Q,\vec P)$ and $\vec p=\vec p(\vec Q,\vec P)$. In particular, the transformation of the canonical variables $(\vec q,\vec p)$ might be given by the unitary transformation above, but this is not required. The generalized Weyl map is then defined by \cite{nuno6}:
\begin{equation}
W^{(q,p)}_{(Q,P)}(\hat A) =  \hbar^N \int d^N\vec x \int d^N\vec y \, e^{-i\vec p(\vec Q,\vec P) \cdot \vec y}
\delta (\vec x-\vec q(\vec Q,\vec P))
<\vec x+\frac{\hbar}{2} \vec y| \hat A |\vec x-\frac{\hbar}{2} \vec y>
\end{equation}
where $|\vec x \pm \frac{\hbar}{2} \vec y>$ are eigenstates of $\hat{\vec q}$.
The new map implements the transformation $(\vec q,\vec p) \to (\vec Q,\vec P)$ as a coordinate transformation in quantum phase space; let $A'(\vec Q,\vec P)=W^{(q,p)}_{(Q,P)}(\hat A)$ and $A(\vec q,\vec p)=W_{(q,p)}(\hat A)$ and we have:
$A'(\vec Q,\vec P)=A(\vec q(\vec Q,\vec P),\vec p(\vec Q,\vec P))$;
though in general it does not preserve the functional form of the star-product and Moyal bracket. Instead it yields the more general covariant star-product and Moyal bracket \cite{nuno6,Flato1}:
\begin{eqnarray}
A'( \vec Q, \vec P) *'_{(Q,P)} B'( \vec Q, \vec P)& = & A'( \vec Q, \vec P) e^{\frac{i \hbar}{2} {\buildrel { \leftarrow}\over\nabla'}_i  J'^{ij}_{(Q,P)}  {\buildrel
{ \rightarrow}\over\nabla'}_j} B' (\vec Q, \vec P) \nonumber \\
\left[ A'(\vec Q,\vec P), B'(\vec Q,\vec P) \right]_{M'_{(Q,P)}} & = & \frac{2}{\hbar} A'(\vec Q,\vec P) \sin \left(\frac{\hbar}{2}
{\buildrel { \leftarrow}\over\nabla'}_i  J'^{ij}_{(Q,P)}  {\buildrel
{ \rightarrow}\over\nabla'}_j \right) B'(\vec Q,\vec P),
\end{eqnarray}
where the covariant derivative is given by (let $O'^i = P_i,  i=1, \cdots ,N$; $O'^i = Q_{i-N},  i=N+1, \cdots, 2N$):
\begin{equation}
\nabla'_i A'  = \partial'_i A', \quad 
\nabla'_i \nabla'_j A' = \partial'_i \partial'_j A' - \Gamma'^k_{ij} \partial'_k A', \quad \partial'_i= \partial /\partial {O'}^{i};
\qquad i,j,k= 1, \cdots, 2N.
\end{equation}
and
\begin{equation}
J'^{ij}_{(Q,P)}(\vec Q,\vec P) = \frac{\partial O'^i}{\partial O^k} \frac{\partial O'^j}{\partial O^l} J^{kl}_{(q,p)}, \quad 
\Gamma'^i_{jk}(\vec Q,\vec P) =  \frac{\partial O'^i}{\partial O^b} \frac{\partial^2 O^b}{\partial O'^j \partial O'^k},
\end{equation}
are the new symplectic structure and Poisson connection associated to the coordinates $(\vec Q,\vec P)$. Notice that in eq.(8) we explicitly took into account the phase space flat structure.

When formulated in terms of these structures Wigner mechanics becomes invariant under the action of general coordinate transformations:
\begin{equation}
A'(\vec Q,\vec P) *^{\prime}_{(Q,P)} B'(\vec Q,\vec P) = A(\vec q(\vec Q,\vec P),\vec p(\vec Q,\vec P))*_{(q,p)} B(\vec q(\vec Q,\vec P),\vec p(\vec Q,\vec P)) \quad \forall_{A,B \in {\cal A}(T^{\ast}M)},
\end{equation}
the covariant generalization of the Moyal and stargenvalue equations reading:
\begin{eqnarray}
& \dot{f'}_W  = [H',f'_W]_{M'_{(Q,P)}} & \nonumber \\
& A' *'_{(Q,P)} g'_a  = g'_a *'_{(Q,P)} A'=  a g'_a, &
\end{eqnarray}
where $f'_W(\vec Q,\vec P;t)=\frac{1}{(2\pi \hbar )^N} W_{(Q,P)}^{(q,p)}(|\psi(t)><\psi(t)|)$ is the covariant Wigner function and $g'_a$ is the left- and right-stargenfunction associated to the eigenvalue $a$.

These equations transform covariantly under arbitrary phase space diffeomorphisms yielding, in any coordinates, identical mathematical solutions and thus identical physical predictions:
\begin{equation}
P(A'(\vec Q,\vec P;t)=a)=  \int d^N\vec Q \int d^N\vec P (\mbox{det} J'^{ij}_{(Q,P)})^{-1/2}
\delta_{*'_{(Q,P)}}(A'(\vec Q,\vec P)-a)  f'_W(\vec Q,\vec P;t),
\end{equation}
where $\delta_{*'_{(Q,P)}}(A'-a)$ is a particular solution of (10), displaying the following explicit form \cite{nuno7,Flato1}:
\begin{equation}
\delta_{*'_{(Q,P)}}(A'(\vec Q,\vec P)-a)=\frac{1}{2\pi}\int dk \; e_{*'_{(Q,P)}}^{ik(A'(\vec Q,\vec P)-a)},
\end{equation}
the $*$-exponential being given by: $e_{*'_{(Q,P)}}^{A'}=\sum_{n=0}^{\infty} \frac{1}{n!} \Omega_n$ where $\Omega_0=1$ and $\Omega_{n+1}=\Omega_n *'_{(Q,P)} A'$.

This concludes our review of the main topics of the covariant formulation of Wigner quantum mechanics.
The reader should refer to \cite{nuno6,Flato1} for more detailed presentations of the theory.

\section{Time dependent canonical transformations}

The aim of this section is to succinctly review some aspects of time dependent canonical transformations in standard operator quantum mechanics.
Let $\hat{\vec A}=\hat{\vec A}(\hat{\vec q},\hat{\vec p},t)$ and $\hat{\vec B}=\hat{\vec B}(\hat{\vec q},\hat{\vec p},t)$ be a new set of fundamental operators $[\hat A_i,\hat B_j]=i\hbar \delta_{ij} , \forall t$. Let $\hat T$ be the generator of the canonical transformation:
\begin{equation}
\frac{\partial}{\partial t} \hat{\vec A}= \frac{1}{i \hbar}[\hat{\vec A},\hat T] \quad \mbox{and} \quad
\frac{\partial}{\partial t} \hat{\vec B}= \frac{1}{i \hbar }[\hat{\vec  B},\hat T],
\end{equation}
and to make it simple let us also impose the initial conditions $\hat{\vec A}(0)= \hat{\vec q}$ and $\hat{\vec B}(0)=\hat{\vec p}$. Then, the unitary transformation reads:
\begin{equation}
\hat{\vec A}= \hat V(t) \hat{\vec q} \,\hat V^{-1}(t) = \hat{\vec A}(\hat{\vec q},\hat{\vec p},t)
\quad \mbox{and} \quad
\hat{\vec B}= \hat V(t) \hat{\vec p} \,\hat V^{-1}(t) = \hat{\vec B}(\hat{\vec q}, \hat{\vec p},t),
\end{equation}
where $\hat V(t)= \exp (\frac{i}{\hbar}\hat Tt)$. The former relations can be immediately inverted: $\hat{\vec q}= \hat V^{-1}(t) \hat{\vec A} \hat V(t)$ and $\hat{\vec p}= \hat V^{-1}(t) \hat{\vec B} \hat V(t)$ and lead straightforwardly to the dynamical equation for a generic observable:
\begin{equation}
\frac{d}{dt} \hat F(\hat{\vec A},\hat{\vec B},t)= \frac{1}{i\hbar } [\hat F, \hat H + \hat T] + \frac{\partial}{\partial t} \hat F(\hat{\vec A},\hat{\vec B},t).
\end{equation}

We now consider the density matrix of the system: $\hat{\rho }(t)= |\psi(t)><\psi (t)|$ where $|\psi(t)> $ is the corresponding quantum state  at the time $t$. In the $\hat q$ representation we have \cite{nuno7}:
\begin{eqnarray}
\hat{\rho }(t) & = & \int d^N\vec q\,' \, d^N\vec q\,''  <\vec q\,''|\psi(t)><\psi(t)|\vec q\,'> |\vec q\,''><\vec q\,'| \nonumber \\
& = &\int d^N\vec q\,' \, d^N\vec q\,'' \; \psi(\vec q\,'',t) \psi^*(\vec q\,',t) e^{-\frac{i}{\hbar} (\vec q\,''-\vec q\,')\cdot \hat{\vec p}} \hat{\Delta} (\hat{\vec q}-\vec q\,') \\
& = & \int d^N\vec q\,'\, d^N\vec q\,'' \; \psi(\vec q\,'',0) \psi^*(\vec q\,',0) e^{-\frac{i}{\hbar} (\vec q\,''-\vec q\,')\cdot  \hat{\vec p}(\hat{\vec q},\hat{\vec p},-t) } \hat{\Delta} (\hat{\vec q}(\hat{\vec q},\hat{\vec p},-t)-\vec q\,') = \hat{\rho }(\hat{\vec q},\hat{\vec p}, t), \nonumber
\end{eqnarray}
where $\hat{\Delta} (\hat{\vec q}-\vec q\,')=\hat{\Delta} (\hat{q}_1- q_1')...
\hat{\Delta} (\hat{q}_N- q_N')$ and $\hat{\Delta} (\hat{q}_i- q_i')=\frac{1}{2\pi}\int dk e^{ik(\hat q_i-q_i')}$. Moreover, $\hat{\vec q}(\hat{\vec q},\hat{\vec p},t)$ and $\hat{\vec p}(\hat{\vec q},\hat{\vec p},t)$ are the Heisenberg time evolutions of the fundamental operators $\hat{\vec q}$ and $\hat{\vec p}$.
The action of the time dependent canonical transformation on the density matrix is now easily implemented,
$
\hat{\rho }(t)= \hat{\rho }(\hat{\vec q}(\hat{\vec A},\hat{\vec B},t),\hat{\vec p}(\hat{\vec A},\hat{\vec B},t),t)= \hat V^{-1}(t) \hat{\rho }(\hat{\vec A},\hat{\vec B},t) \hat V(t) = \hat{\rho }'(\hat{\vec A},\hat{\vec B},t)
$, 
from where it follows that in the $\hat A$ representation:
\begin{eqnarray}
\frac{\partial}{\partial t} \hat{\rho }'(\hat{\vec A},\hat{\vec B},t) & = & \frac{\partial \hat V^{-1}(t)}{\partial t} \hat{\rho} \hat V(t)+ \hat V^{-1}(t) \left(\frac{\partial}{\partial t} \hat{\rho}(\hat{\vec A},\hat{\vec B}, t) \right)\hat V(t)+
\hat V^{-1}(t) \hat{\rho} \frac{\partial \hat V(t)}{\partial t} = \nonumber \\
&=& \frac{1}{i\hbar }[\hat T, \hat{\rho}'] + \frac{1}{i\hbar }\hat V^{-1} [\hat H, \hat{\rho}] \hat V = \frac{1}{i\hbar }[\hat T+\hat H', \hat{\rho}'],
\end{eqnarray}
where $\hat H'(\hat{\vec A},\hat{\vec B},t)=\hat H (\hat{\vec q}(\hat{\vec A},\hat{\vec B},t) ,\hat{\vec p}(\hat{\vec A},\hat{\vec B},t))=\hat V^{-1}(t) \hat H(\hat{\vec A},\hat{\vec B}) \hat V (t)$.

\section{Time dependent transformations in phase space quantum mechanics}

In this section we study the action of time dependent transformations in phase space quantum mechanics.
We consider an arbitrary $N$ dimensional quantum system with Hamiltonian $\hat H$ and described by the wave function $\psi(t)$. As we have seen, the original Weyl transform $W_{( q, p)}$ yields the standard Wigner formulation of the system. The time evolution of the Wigner function $f_W(\vec q,\vec p,t)=\frac{1}{(2\pi \hbar )^N}W_{(q,p)}(|\psi(t)><\psi(t)|)$ is dictated by the standard Moyal equation where the Moyal bracket and the starproduct are given by eq.(2).
We then consider two different phase space implementations of a time dependent operator transformation.

(1) {\it Unitary time dependent transformations and the map $W_{(A,B)}$.}\\
At the quantum operator level we introduce the unitary transformation $(\hat{\vec q},\hat{\vec p}) \longrightarrow
(\hat{\vec A},\hat{\vec B})$ defined by eqs.(13,14). The new variables $(\hat{\vec A},\hat{\vec B})$ satisfy the Heisenberg commutation relations and thus a new Weyl map $W_{(A,B)}$ can be constructed. It displays the standard functional structure given by eq.(1) and
yields a starproduct $*_{(A,B)}$ and Moyal bracket $[,]_{M_{(A,B)}}$ also displaying the non-covariant functional form eq.(2). The time evolution of the new
Wigner function:
\begin{equation}
f'_W(\vec A,\vec B,t)=\frac{1}{(2\pi \hbar )^N} W_{(A,B)} \left(\hat{\rho }'(\hat{\vec A},\hat{\vec B},t)\right)=V^{-1}(t)*_{(A,B)}f_W(\vec A,\vec B,t)*_{(A,B)} V(t),
\end{equation}
where $V(t)=W_{(A,B)} (\hat V)$, reads:
\begin{equation}
\frac{\partial}{\partial t} f_W'(\vec A,\vec B,t)=[H'+T,f'_W]_{M_{(A,B)}},
\end{equation}
and is just the $(A,B)$-Weyl transform of eq.(17). Notice that just like in the time independent case the unitary transformation $(\vec q,\vec p) \to (\vec A,\vec B)$ does not, in general act as a coordinate transformation: $ f_W'(\vec A,\vec B,t) \not= f_W(\vec q(\vec A,\vec B,t),\vec p(\vec A,\vec B,t),t)$.

(2) {\it Coordinate transformations and the map $W_{(A,B)}^{(q,p)}$}.\\
We follow the steps of the time independent case and introduce a time dependent phase space diffeomorphism $(\vec q,\vec p) \longrightarrow (\vec A,\vec B)$ defined in generic terms by $\vec q=\vec q(\vec A,\vec B,t)$; $\vec p=\vec p(\vec A,\vec B,t)$. This coordinate transformation is not required to be a symplectomorphism (i.e. to preserve the Poisson bracket) nor to preserve the Moyal bracket between the fundamental variables (i.e. to satisfy $[q_i(\vec A,\vec B,t),p_j(\vec A,\vec B,t)]_{M_{(A,B)}}=\delta_{ij}$ and $[q_i(\vec A,\vec B,t),q_j(\vec A,\vec B,t)]_{M_{(A,B)}}=[p_i(\vec A,\vec B,t),p_j(\vec A,\vec B,t)]_{M_{(A,B)}}=0$ for all $i,j=1...N$).

We then define the time dependent generalized Weyl transform in the variables $(\vec A, \vec B)$:
\begin{equation}
W_{(A,B)}^{(q,p)}: \hat{\cal A}({\cal H}) \longrightarrow {\cal A}(T^{\ast}M); \quad
\hat F \longrightarrow F'(\vec A,\vec B,t)=
W_{(A,B)}^{(q,p)}(\hat F)=\left. W_{(q,p)}(\hat F)\right|_{\vec q=\vec q(\vec A,\vec B,t) \wedge \vec p=\vec p(\vec A,\vec B,t)}.
\end{equation}
The explicit form of $W_{(A,B)}^{(q,p)}$ is given by the trivial time dependent generalization of eq.(5):
\begin{equation}
W^{(q,p)}_{(A,B)}(\hat F)= \hbar^N \int d^N\vec x \int d^N\vec y \; e^{-i\vec p(\vec A,\vec B,t) \cdot \vec y}
\delta (\vec x-\vec q(\vec A,\vec B,t))
<\vec x+\frac{\hbar}{2} \vec y| \hat F |\vec x-\frac{\hbar}{2} \vec y>,
\end{equation}
from which follows the covariant {\it time dependent} $*$-product and Moyal bracket:
\begin{eqnarray}
W_{(A,B)}^{(q,p)}(\hat F \hat G) & = & F'(\vec A,\vec B,t) *^{\prime}_{(A,B)} G'(\vec A,\vec B,t) \nonumber \\
& = & F(\vec q(\vec A,\vec B,t),\vec p(\vec A,\vec B,t))*_{(q,p)} G(\vec q(\vec A,\vec B,t),\vec p(\vec A,\vec B,t)),
\end{eqnarray}
and $[F',G']_{M'_{(A,B)}}=\frac{1}{i\hbar} \left( F'*'_{(A,B)}G'-G'*'_{(A,B)}F' \right)$
where $F'(\vec A,\vec B,t)=W_{(A,B)}^{(q,p)}(\hat F)$, $ F(\vec q,\vec p,t)=W_{(q,p)}(\hat F)$ and likewise for $G$ and $G'$. The two algebraic structures display the functional form given by eqs.(6-8) with the obvious inclusion of an explicit time dependence.

The dynamical structure of the theory displays more significant corrections.
Let $f_W'(\vec A,\vec B,t)=\frac{1}{(2\pi \hbar )^N}W_{(A,B)}^{(q,p)}(|\psi(t)><\psi(t)|)$ be the covariant Wigner function. It satisfies
$f_W'(\vec A,\vec B,t)=f_W(\vec q(\vec A,\vec B,t),\vec p(\vec A,\vec B,t),t)$ and thus:
\begin{eqnarray}
& & \frac{\partial }{\partial t} f_W'(\vec A,\vec B,t)  = \left. \left( \frac{\partial }{\partial t_1}+\frac{\partial }{\partial t_2} \right) f_W(\vec q(\vec A,\vec B,t_1),\vec p(\vec A,\vec B,t_1),t_2) \right|_{t_1=t_2=t} \nonumber \\
& = & [H(\vec q(\vec A,\vec B,t),\vec p(\vec A,\vec B,t),t) ,f_W(\vec q(\vec A,\vec B,t),\vec p(\vec A,\vec B,t),t)]_{M_{(q,p)}} + \left. \frac{\partial f_W}{\partial \vec q}\cdot \frac{\partial \vec q}{\partial t_1}+\frac{\partial f_W}{\partial \vec p} \cdot \frac{\partial \vec p }{\partial t_1} \right|_{t_1=t} \nonumber \\
&=& [H'(\vec A,\vec B,t),f_W'(\vec A,\vec B,t)]_{M'_{(A,B)}} + \left( \frac{\partial f'_W}{\partial A_i}\frac{\partial A_i}{\partial q_j}+\frac{\partial f'_W}{\partial B_i}\frac{\partial B_i}{\partial q_j} \right) \frac{\partial q_j}{\partial t}
+ \left(\frac{\partial f'_W}{\partial A_i}\frac{\partial A_i}{\partial p_j} + \frac{\partial f'_W}{\partial B_i}\frac{\partial B_i}{\partial p_j} \right) \frac{\partial p_j}{\partial t} \nonumber \\
& = & [H'(\vec A,\vec B,t),f_W'(\vec A,\vec B,t)]_{M'_{(A,B)}} +
\frac{\partial f'_W}{\partial A_i} \left( \frac{\partial A_i}{\partial q_j}\frac{\partial q_j}{\partial t}+\frac{\partial A_i}{\partial p_j}\frac{\partial p_j}{\partial t} \right) + \frac{\partial f_W'}{\partial B_i} \left( \frac{\partial  B_i}{\partial q_j}\frac{\partial q_j}{\partial t}+\frac{\partial B_i}{\partial p_j}\frac{\partial p_j}{\partial t}\right) \nonumber \\
&=& [H'(\vec A,\vec B,t),f_W'(\vec A,\vec B,t)]_{M'_{(A,B)}} - \frac{\partial f'_W}{\partial \vec A} \cdot \frac{\partial \vec A}{\partial t} - \frac{\partial f'_W}{\partial \vec B} \cdot \frac{\partial \vec B}{\partial t},
\end{eqnarray}
where in the last step we used the fact that $\vec A=\vec A(\vec q(\vec A,\vec B,t),\vec p(\vec A,\vec B,t),t)$ and likewise for $\vec B$. Further, contraction over repeated indices is understood, i.e. $ \frac{\partial f_W}{\partial q_i}\frac{\partial q_i}{\partial t_1} = \frac{\partial f_W}{\partial \vec q}\cdot \frac{\partial \vec q}{\partial t_1}=  \sum_{i=1}^{N} \frac{\partial f_W}{\partial q_i}\frac{\partial q_i}{\partial t_1}$.
Equation (23) constitutes a generalization of the Moyal covariant equation (10) and renders the dynamics of the Wigner function fully invariant under the action of general time dependent phase space diffeomorphisms.

Let us then consider several particular cases in more detail:\\
a) If the transformation $(\vec q,\vec p) \to (\vec A,\vec B)$ is time independent then eq.(23) reduces to the standard covariant Moyal equation (10).\\
b) If, on the other hand, it is unitary, i.e. if $\vec A(\vec q,\vec p,t)=V(t)*_{(q,p)}\vec q*_{(q,p)}V^{-1}(t)$
and $\vec B(\vec q,\vec p,t)=V(t)*_{(q,p)}\vec p*_{(q,p)}V^{-1}(t)$ satisfying $\frac{\partial \vec A}{\partial t}=[\vec A,T]_{M_{(q,p)}}$ and $\frac{\partial \vec B}{\partial t}=[\vec B,T]_{M_{(q,p)}}$ with initial conditions $
\vec A(\vec q,\vec p,0)=\vec q$ and $\vec B(\vec q,\vec p,0)=\vec p$ then in eq.(23) we have:
{\begin{eqnarray}
\frac{\partial \vec A}{\partial t} & = & \left.\frac{\partial }{\partial t}\vec A(\vec q,\vec p,t)\right|_{\vec q=\vec q(\vec A,\vec B,t) \wedge \vec p=\vec p(\vec A,\vec B,t)} \\
& = & [\vec A(\vec q(\vec A,\vec B,t),\vec p(\vec A,\vec B,t),t),T(\vec q(\vec A,\vec B,t),\vec p(\vec A,\vec B,t))]_{M_{(q,p)}}=[\vec A,T'(\vec A,\vec B,t)]_{M'_{(A,B)}} \nonumber
\end{eqnarray}
where $T'(\vec A,\vec B,t)=W_{(A,B)}^{(q,p)}(\hat T)$. An equivalent result is valid for $\frac{\partial \vec B}{\partial t}$. Substituting these results in eq.(23) we get:
\begin{equation}
\frac{\partial }{\partial t} f_W'(\vec A,\vec B,t) =[H'(\vec A,\vec B,t),f_W'(\vec A,\vec B,t)]_{M'_{(A,B)}} + \frac{\partial f_W'}{\partial \vec A}\cdot [T',\vec A]_{M'_{(A,B)}}+  \frac{\partial f_W'}{\partial \vec B}\cdot [T',\vec B]_{M'_{(A,B)}}
\end{equation}
c) Finally, we consider the case where the transformation $(\vec q,\vec p) \to (\vec A,\vec B)$ is a symplectomorphism.
Let $T$ be the generator. Then $\vec A(\vec q,\vec p,t)$ satisfies $\frac{\partial \vec A}{\partial t}=\{\vec A,T\}_{(q,p)}=\{\vec A,T\}_{(A,B)}$ and likewise for $\vec B$. Hence, eq.(23) reduces to:
\begin{eqnarray}
\frac{\partial }{\partial t} f_W'(\vec A,\vec B,t) & = & [H'(\vec A,\vec B,t),f_W'(\vec A,\vec B,t)]_{M'_{(A,B)}} - \frac{\partial f'_W}{\partial \vec A} \cdot  \{\vec A,T\}_{(A,B)}
- \frac{\partial f'_W}{\partial \vec B} \cdot \{\vec B,T\}_{(A,B)} \nonumber \\
&=& [H'(\vec A,\vec B,t),f_W'(\vec A,\vec B,t)]_{M'_{(A,B)}} + \{T,f'_W\}_{(A,B)}
\end{eqnarray}

To finish this section let us study the $^{(q,p)}_{(A,B)}$-representation of a general stargenfunction.
Let $\hat F$ be a generic operator and $F'(\vec A,\vec B,t)= W^{(q,p)}_{(A,B)} (\hat F)$. The $*$-genvalue equation in the $^{(q,p)}_{(A,B)}$-representation is then written:
\begin{equation}
F'(\vec A,\vec B,t)*'_{(A,B)} g'_a(\vec A,\vec B,t)= g'_a(\vec A,\vec B,t)*'_{(A,B)} F'(\vec A,\vec B,t)=
ag'_a(\vec A,\vec B,t)
\end{equation}
and displays the solution:
\begin{eqnarray}
g'_a(\vec A,\vec B,t) & = & \delta_{*'_{(A,B)}} [F'(\vec A,\vec B,t)-a] =\frac{1}{2\pi} \int dk e_{*'_{(A,B)}}^{ik[F'(\vec A,\vec B,t)-a]} \nonumber \\
&=& \frac{1}{2\pi}\int dk e_{*_{(q,p)}}^{ik[F(\vec q(\vec A,\vec B,t),\vec p(\vec A,\vec B,t),t)-a]}
\end{eqnarray}
where $F(\vec q,\vec p,t)=W_{(q,p)} (\hat F)$. Further if $g_a(\vec q,\vec p,t)$ is such that $F*_{(q,p)} g_a=g_a*_{(q,p)}F=a g_a$ then $g'_a(\vec A,\vec B,t)=g_a(\vec q(\vec A,\vec B,t),\vec p(\vec A,\vec B,t),t)$, a result that follows immediately from eqs.(22,28). Therefore the stargenvalue equation transforms covariantly under the action of general time dependent coordinate transformations.

As an illustrative example let us consider the one-dimensional simple case
$\hat F= \hat q \Longrightarrow W^{(q,p)}_{(A,B)} (\hat q) = q(A,B,t)$. The solution of the $*$-genvalue equation (27) is then:
\begin{eqnarray}
g'_a(A,B,t) & = & \frac{1}{2\pi}\int dk e_{*'_{(A,B)}}^{ik[q(A,B,t)-a]}
= \frac{1}{2\pi} \int dk e_{*_{(q,p)}}^{ik[q(A,B,t)-a]} \nonumber \\
& = & \frac{1}{2\pi} \int dk e^{ik[q(A,B,t)-a]}
= \delta [q(A,B,t)-a],
\end{eqnarray}
and displays the time evolution:
\begin{equation}
\frac{\partial }{\partial t} g'_a(A,B,t) =  \frac{\partial \delta}{\partial q} [q(A,B,t)-a]\frac{\partial q}{\partial t}(A,B,t) .
\end{equation}
Furthermore, if the transformation $(q,p) \to (A,B)$ is symplectic with generator $T$ then $\frac{\partial q}{\partial t}=\{q,-T\}_{(A,B)}$ and,
\begin{equation}
\frac{\partial }{\partial t} g'_a(A,B,t) = -\{g_a',T\}_{(A,B)}.
\end{equation}

\section{The causal representation}

As an application of the formalism let us consider a finite dimensional dynamical system described by a generic Hamiltonian $\hat H$ and use the generalized time dependent Weyl map to derive in a systematic way: 1) the Schr\"odinger and 2) the Heisenberg phase space pictures and 3) a new phase space representation where the Wigner function displays a fully classical time evolution.

To begin with, we introduce the time dependent unitary transformation generated by $\hat T=-\hat H$. A new set of fundamental operators is given by $\hat{\vec A}=\hat{\vec A}(\hat{\vec q},\hat{\vec p},t)$ and $\hat{\vec B}=\hat{\vec B}(\hat{\vec q},\hat{\vec p},t)$, solutions of eq.(13) and satisfying the initial conditions $\hat{\vec A}(0)=\hat{\vec q}$ and $\hat{\vec B}(0)=\hat{\vec p}$. Given the relation between $\hat H$ and $\hat T$ they also satisfy:
\begin{equation}
\hat{\vec A}=\hat{\vec A}(\hat{\vec q},\hat{\vec p},t)= \hat{\vec q}(\hat{\vec q},\hat{\vec p}, -t) \quad \mbox{and} \quad
\hat{\vec B}=\hat{\vec B}(\hat{\vec q},\hat{\vec p},t)= \hat{\vec p}(\hat{\vec q},\hat{\vec p}, -t)
\end{equation}
where $\hat{\vec q}(\hat{\vec q},\hat{\vec p},t)$ and $\hat{\vec p}(\hat{\vec q},\hat{\vec p}, t) $ are the Heisenberg time evolution of the fundamental operators $\hat{\vec q}$ and $\hat{\vec p}$.
From (32) we define a new set of phase space coordinates $\vec A=\vec A_M(\vec q,\vec p,t)=W_{(q,p)}(\hat A)$ and $\vec B=\vec B_M(\vec q,\vec p,t)=W_{(q,p)}(\hat B)$ satisfying the Moyal equations:
\begin{equation}
\frac{\partial \vec A_M}{\partial t}=[\vec A_M,T]_{M_{(q,p)}} \quad , \quad
\frac{\partial \vec B_M}{\partial t}=[\vec B_M,T]_{M_{(q,p)}},\quad T=W_{(q,p)}(\hat T)
\end{equation}
and the initial conditions: $\vec A_M(\vec q,\vec p, 0)=\vec q$ and $\vec B_M(\vec q,\vec p,0)=\vec p$.
The subscript $M$ indicates that the {\it functions} $A_M$ and $B_M$ obey the Moyal equations. Intuitively, they can be seen as the quantum phase space histories of the system.

A second set of (purely classical) phase space coordinates will also be required: $\vec Q,\vec P$ are given by $\vec Q=\vec Q(\vec A,\vec B,t)$ and $\vec P=\vec P(\vec A,\vec B,t)$ and satisfy the classical Hamiltonian equations:
\begin{equation}
\frac{\partial \vec Q}{\partial t}=\{\vec Q,-T(\vec A,\vec B)\}_{(A,B)} , \quad
\frac{\partial \vec P}{\partial t}=\{\vec P,-T(\vec A,\vec B)\}_{(A,B)}, \quad T(\vec A,\vec B)=W_{(A,B)}(\hat T)
\end{equation}
and the initial conditions $\vec Q(\vec A,\vec B,0)=\vec A$, $\vec P(\vec A,\vec B,0)=\vec B$. Notice that $T(\vec A,\vec B)$ displays the same functional form as $T(\vec q,\vec p)=W_{(q,p)}(\hat T)$. Solving the algebraic equations $\vec Q=\vec Q(\vec A,\vec B,t)$, $\vec P=\vec P(\vec A,\vec B,t)$ with respect to $\vec A,\vec B$ we get:
$\vec A=\vec A_C(\vec Q,\vec P,t)$ and $\vec B=\vec B_C(\vec Q,\vec P,t)$ where the subscript $C$ indicates that $\vec A_C$ and $\vec B_C$ are the solutions of the classical Hamilton's equations:
\begin{equation}
\frac{\partial \vec A_C}{\partial t}=\{\vec A_C,T(\vec Q,\vec P)\}_{(Q,P)} \quad , \quad
\frac{\partial \vec B_C}{\partial t}=\{\vec B_C,T(\vec Q,\vec P)\}_{(Q,P)}
\end{equation}
where $T(\vec Q,\vec P)=T(\vec A_C(\vec Q,\vec P,t),\vec B_C(\vec Q,\vec P,t))$ displays the same functional form as $T(\vec A,\vec B)$ (notice that $T$ is the generator of the canonical transformation).
Also notice that in general $\vec q(\vec A,\vec B,t) \not= \vec Q(\vec A,\vec B,t)$ and $\vec p(\vec A,\vec B,t) \not= \vec P(\vec A,\vec B,t)$ (the exceptions happen for $T$ quadratic in the phase space variables). While the variables $(\vec q, \vec p)$ describe the {\it quantum} phase space time evolution, the variables $(\vec Q,\vec P)$ describe the {\it classical} phase space trajectories.
The transformation $(\vec A,\vec B) \to (\vec Q,\vec P)$ is a phase space symplectomorphism exclusively defined at the classical level, i.e. it is not (and unlike $(\vec q,\vec p)\to (\vec A,\vec B)$ it could not be) inherited from a quantum operator transformation.

Finally, from eqs.(16,32) the density matrix $\hat{\rho }(t) = |\psi(t)><\psi(t)|$ admits the expansions:
\begin{eqnarray}
\hat{\rho }(t) & = &
\int d^N\vec q\,' d^N\vec q\, '' \psi(\vec q\,'',0) \psi^*(\vec q\,',0) e^{-\frac{i}{\hbar} (\vec q\,''-\vec q\,')\cdot  \hat{\vec p}(\hat{\vec q},\hat{\vec p},-t) } \hat{\Delta} [\hat{\vec q}(\hat{\vec q},\hat{\vec p},-t)-\vec q\,'] = \hat{\rho }(\hat{\vec q},\hat{\vec p}, t), \nonumber \\
\hat{\rho }(t) & = &
\int d^N\vec q\,'d^N\vec q\,'' \psi(\vec q\,'',0) \psi^*(\vec q\,',0) e^{-\frac{i}{\hbar} (\vec q\,''-\vec q\,') \cdot \hat{\vec B}} \hat{\Delta} (\hat{\vec A}-\vec q\,') = \hat{\rho }(\hat{\vec A},\hat{\vec B}, 0)
\end{eqnarray}

With these preliminaries settled down we address the derivation of the three phase space pictures:

(1) {\it Schr\"odinger picture and the map $W_{(q,p)}$}.\\
From the first expansion for the density matrix (36) we immediately get:
\begin{eqnarray}
& & f_W(\vec q,\vec p,t)  =  \frac{1}{(2\pi \hbar)^N} W_{(q,p)} [\hat{\rho }(t)]  \\
& = & \frac{1}{(2\pi \hbar)^N}
\int d^N\vec q\,'d^N\vec q\,'' \psi(\vec q\,'',0) \psi^*(\vec q\,',0) e_{*(q,p)}^{-\frac{i}{\hbar} (\vec q\,''-\vec q\,') \cdot  \vec p(\vec q,\vec p ,-t) } *_{(q,p)} {\delta}_{*(q,p)} [\vec q(\vec q,\vec p,-t)-\vec q\,'] \nonumber
\end{eqnarray}
and the time evolution of the Wigner function obeys the standard Moyal equation: $\frac{\partial }{\partial t} f_W=[H,f_W]_{M_{(q,p)}}$. We also have: $W_{(q,p)}(|\vec q_0><\vec q_0|)=\delta_*(\vec q-\vec q_0)= \delta(\vec q-\vec q_0)$ and so $\frac{\partial }{\partial t} \delta (\vec q-\vec q_0) =0$ and likewise for $\vec p$, i.e. the dynamics is cast in the Schr\"odinger picture.

(2) {\it Heisenberg picture and the map $W_{(A,B)}$}.\\
From the second expansion in (36) we get:
\begin{eqnarray}
&& \frac{1}{(2\pi \hbar)^N} W_{(A,B)} [\hat{\rho }(t)]  =  \frac{1}{(2\pi \hbar)^N} W_{(A,B)} [\hat{\rho }(\vec A,\vec B,0)]\\
& = &
\frac{1}{(2\pi \hbar)^N} \int d^N \vec q\, 'd^N \vec q\, '' \psi(\vec q\,'',0) \psi^*(\vec q\,',0) e_{*(A,B)}^{-\frac{i}{\hbar} (\vec q\,''-\vec q\,') \cdot \vec B } *_{(A,B)} {\delta^N}_{*(A,B)} (\vec A-\vec q\,') \nonumber
\end{eqnarray}
Let us check explicit that the previous formula yields the standard (in this case time independent) expression of the Wigner function. The same derivation would also apply to eq.(37). We start by considering the term evolving starproducts in more detail:
\begin{eqnarray}
&& e_{*(A,B)}^{-\frac{i}{\hbar} (\vec q\,''-\vec q\,') \cdot \vec B } *_{(A,B)} {\delta^N}_{*(A,B)} (\vec A-\vec q\,') \quad = \quad e^{-\frac{i}{\hbar} (\vec q\,''-\vec q\,') \cdot \vec B } *_{(A,B)} {\delta^N} (\vec A-\vec q\,') \nonumber \\
& = & \frac{1}{(2\pi)^N} \sum_{n=0}^{+\infty} \frac{1}{n!} \left(-\frac{i\hbar}{2} \right)^n e^{-\frac{i}{\hbar} (\vec q\,''-\vec q\,') \cdot \vec B } \left[-\frac{i}{\hbar} (\vec q\,''-\vec q\,') \cdot \frac{\partial}{\partial \vec A} \right]^n \int d^N \vec k \, e^{i\vec k \cdot (\vec A-\vec q\,')} \nonumber \\
&=& \frac{1}{(2\pi)^N} e^{-\frac{i}{\hbar} (\vec q\,''-\vec q\,') \cdot \vec B }
\int d^N \vec k \sum_{n=0}^{+\infty} \frac{1}{n!} \left(-\frac{i}{2} \vec k \cdot (\vec q\,''-\vec q\,') \right)^n    e^{i\vec k \cdot (\vec A-\vec q\,')}  \\
&=&  e^{-\frac{i}{\hbar} (\vec q\,''-\vec q\,') \cdot \vec B }
\frac{1}{(2\pi)^N} \int d^N \vec k \, e^{i\vec k \cdot (\vec A-\vec q\,'+\frac{\vec q\,'}{2}-\frac{\vec q\,''}{2})} \quad
=  \quad e^{-\frac{i}{\hbar} (\vec q\,''-\vec q\,') \cdot \vec B }
\delta^N \left(\vec A-\frac{\vec q\,'}{2}-\frac{\vec q\,''}{2}\right). \nonumber
\end{eqnarray}
Substituting this expression in (38) we get:
\begin{eqnarray}
&& \frac{1}{(2\pi \hbar)^N} W_{(A,B)} [\hat{\rho }(t)] = \nonumber \\
& = & \frac{1}{(2\pi \hbar)^N}\int d^N\vec q\,'d^N\vec q\,'' \psi(\vec q\,'',0) \psi^*(\vec q\,',0) 2^N {\delta^N} (2 \vec A-\vec q\,'- \vec q\,'')
e^{-\frac{2i}{\hbar} (\vec A-\vec q\,')  \cdot \vec B } \nonumber \\
&=& \frac{1}{(\pi \hbar)^N}\int d^N\vec q\,' \psi(2\vec A-\vec q\,',0) \psi^*(\vec q\,',0)
e^{-\frac{2i}{\hbar} (\vec A-\vec q\,')\cdot  \vec B } \nonumber \\
& = & \frac{1}{(\pi \hbar)^N}\int d^N\vec y \psi(\vec A+\vec y,0) \psi^*(\vec A-\vec y,0)
e^{-\frac{2i}{\hbar} \vec y \cdot \vec B } = f_W(\vec A,\vec B,0),
\end{eqnarray}
where in the last step we made $\vec y=\vec A-\vec q\,'$. We indeed recovered the standard definition of the Wigner function. Moreover we see that $\frac{\partial }{\partial t} f_W=0$, as it should and in perfect agreement with eq.(19) taking into account that $T=-H$.
On the other hand we also have:
\begin{eqnarray}
g_{\vec q_0}(\vec A,\vec B,t) & = & W_{(A,B)} (|\vec q_0><\vec q_0|)  =  \delta^N_{*_{(A,B)}} [\vec q(\vec A,\vec B,t)-\vec q_0] \nonumber \\
& = & \frac{1}{(2\pi)^N} \int d^N \vec k \, e_{*_{(A,B)}}^{i\vec k \cdot (\vec q(\vec A,\vec B,t)-\vec q_0)},
\end{eqnarray}
where $\vec q=\vec q(\vec A,\vec B,t)$ is the solution with respect to $\vec q$ of the algebraic equations $\vec A=\vec A_M(\vec q,\vec p,t)$, $\vec B=\vec B_M(\vec q,\vec p,t)$ defined in eq.(33). It also follows from eq.(33) that $\vec q(\vec A,\vec B,t)$ satisfies $\frac{\partial \vec q}{\partial t}=[\vec q,-T]_{M_{(A,B)}}$. Therefore:
\begin{equation}
\frac{\partial }{\partial t} g_{\vec q_0}(\vec A,\vec B,t) =  [g_{\vec q_0}(\vec A,\vec B,t),H]_{M_{(A,B)}},
\end{equation}
and we obtained the phase space Heisenberg picture.

(3) {\it Causal picture and the map $W_{(Q,P)}^{(A,B)}$}.\\
We finally consider the action of the map $W_{(Q,P)}^{(A,B)}$ on the second expansion for the density matrix (36). From eq.(40) it follows that (notice that $W_{(Q,P)}^{(A,B)}=W_{(A,B)}$):
\begin{eqnarray}
& & f'_W(\vec Q,\vec P,t)=\frac{1}{(2\pi \hbar)^N} W_{(Q,P)}^{(A,B)} [\hat{\rho }(t)] \nonumber \\
& = & \frac{1}{(2\pi \hbar)^N} \int d^N\vec q\,'d^N\vec q\,'' \psi(\vec q\,'',0) \psi^*(\vec q\,',0) e_{*'(Q,P)}^{-\frac{i}{\hbar} (\vec q\,''-\vec q\,') \cdot \vec B_C(\vec Q,\vec P,t) } *'_{(Q,P)} {\delta^N_{*'(Q,P)} } [\vec A_C(\vec Q,\vec P,t)-\vec q\,'] \nonumber \\
& = & \frac{1}{(2\pi \hbar)^N} \int d^N\vec q\,'d^N\vec q\,'' \psi(\vec q\,'',0) \psi^*(\vec q\,',0) e_{*(A,B)}^{-\frac{i}{\hbar} (\vec q\,''-\vec q\,') \cdot \vec B_C(\vec Q,\vec P,t) } *_{(A,B)} {\delta^N_{*(A,B)} } [\vec A_C(\vec Q,\vec P,t)-\vec q\,'] \nonumber \\
& = & f_W(\vec A_C(\vec Q,\vec P,t),\vec B_C(\vec Q,\vec P,t),0)=f_W(\vec Q(\vec Q,\vec P,-t),\vec P(\vec Q,\vec P,-t),0)
\end{eqnarray}
where $\vec Q(\vec Q,\vec P,t)=\vec A_C(\vec Q,\vec P,-t)$ and $\vec P(\vec Q,\vec P,t)=\vec B_C(\vec Q,\vec P,-t)$ are the classical time evolution of the canonical variables $(\vec Q,\vec P)$ (cf.(35)):
\begin{equation}
\dot{\vec Q}=\{\vec Q,-T(\vec Q,\vec P)\}_{(Q,P)} \quad \mbox{and} \quad \dot{\vec P}=\{\vec P,-T(\vec Q,\vec P)\}_{(Q,P)}
\end{equation}
associated with the Hamiltonian $H(\vec Q,\vec P)=-T(\vec Q,\vec P)= W_{(A,B)}(\hat H)|_{\vec A=\vec Q \wedge \vec B=\vec P}$. Hence, the time evolution of the Wigner function is given by:
\begin{equation}
\frac{\partial }{\partial t} f'_W(\vec Q,\vec P,t)=\{f'_W,T\}_{(Q,P)}
=\{H,f'_W\}_{(Q,P)}
\end{equation}
On the other hand for the $*$-genfunctions $W_{(Q,P)}^{(A,B)} (|\vec q_0><\vec q_0|) $ we have from eq.(41):
\begin{eqnarray}
g_{\vec q_0}'(\vec Q,\vec P,t)& = & \delta^N_{*'_{(Q,P)}}(\vec q-\vec q_0)=\left. \delta^N_{*_{(A,B)}}(\vec q(\vec A,\vec B,t)-\vec q_0) \right|_{\vec A=\vec A_C(\vec Q,\vec P,t) \wedge \vec B=\vec B_C(\vec Q,\vec P,t) } \nonumber \\
& = & g_{\vec q_0} (\vec A_C(\vec Q,\vec P,t),\vec B_C(\vec Q,\vec P,t),t),
\end{eqnarray}
and so (cf.(35,42)):
\begin{eqnarray}
\frac{\partial }{\partial t} g_{\vec q_0}'(\vec Q,\vec P,t)& = &
\left. \left\{\frac{\partial }{\partial t_1}+\frac{\partial }{\partial t_2} \right\} g_{\vec q_0} (\vec A_C(\vec Q,\vec P,t_2),\vec B_C(\vec Q,\vec P,t_2),t_1) \right|_{t_1=t_2=t} \nonumber \\
& = &
[g_{\vec q_0}',H]_{M'_{(Q,P)}}-\{g_{\vec q_0}',H\}_{(Q,P)}.
\end{eqnarray}
In this representation both the Wigner function and the stargenfunctions evolve in time, the Wigner function displaying a fully classical evolution. In particular if the Wigner function is positive defined at the initial time it will remain so for all times. We conclude that the source of the quantum behavior has been completely removed from the distributional sector of the theory and is now exclusively placed on the observables (stargenfunctions) sector.

\section{Example}

To illustrate our previous results let us consider a two particle system described by the Hamiltonian:
\begin{equation}
\hat H= \frac{\hat p_1^2}{2M} + \frac{\hat p_2^2}{2m} + k \hat q_1\hat p_2^2
\end{equation}
where $(\hat q_1,\hat p_1)$ are the canonical variables of the particle of mass $M$, $(\hat q_2,\hat p_2)$ those of the particle of mass $m$ and $k$ is a coupling constant.

The Weyl map $W_{(q,p)}$ yields the $(q,p)$-Hamiltonian symbol:
\begin{equation}
H= W_{(q,p)}(\hat H)=\frac{ p_1^2}{2M} + \frac{ p_2^2}{2m} + k q_1 p_2^2
\end{equation}
and the Moyal equations $\dot z=[z,H]_{M_{(q,p)}}$ for the fundamental variables $z=q_1,q_2,p_1$ or $p_2$. These display the solutions:
\begin{equation}
\left\{ \begin{array}{lll}
q_1(t) & = & q_1(0) + \frac{p_1(0)}{M} t - \frac{k}{2M} p_2(0)^2t^2 \\
p_1(t) & = & p_1(0)-kp_2(0)^2t \\
q_2(t) & = & q_2(0) + \left\{\frac{p_2(0)}{m} + 2kq_1(0)p_2(0) \right\}t + \frac{k}{M} p_1(0)p_2(0)t^2 - \frac{k^2}{3M}p_2(0)^3t^3 \\
p_2(t) & = & p_2(0)
\end{array} \right.
\end{equation}
which coincide exactly with the classical time evolution, i.e. with the solutions of the classical Hamiltonian equations for the classical Hamiltonian (49). This property is not shared by the Wigner function, its time evolution satisfying the equation:
\begin{equation}
\frac{\partial f_W}{\partial t}=[H,f_W]_{M_{(q,p)}} \Longleftrightarrow \frac{\partial f_W}{\partial t}=\{H,f_W\}_{{(q,p)}}
+\frac{\hbar^2}{24} \left[2\left\{2kp_2,\frac{\partial^2 f_W}{\partial q_2 \partial p_1}\right\}_{(q,p)} -
\left\{2kq_1,\frac{\partial^2 f_W}{\partial q_2^2}\right\}_{(q,p)} \right]
\end{equation}
which is obviously not of the form of the Liouville equation. Consequently, the Wigner function does not satisfy:
$f_W(\vec q,\vec p,t)=f_W(\vec q(-t),\vec p(-t),0)$ with $\vec q(t),\vec p(t)$ given by eq.(50) and $\vec q=(q_1,q_2)$, $\vec p=(p_1,p_2)$, i.e. it does not display a classical causal structure.

We now introduce a new set of fundamental operators:
\begin{equation}
\left\{ \begin{array}{lll}
\hat A_1 & = & \hat q_1 - \frac{\hat p_1}{M} t - \frac{k}{2M} \hat p_2^2t^2 \\
\hat B_1 & = & \hat p_1+k\hat p_2^2t \\
\hat A_2 & = & \hat q_2 - \left\{\frac{\hat p_2}{m} + 2k\hat q_1\hat p_2 \right\}t + \frac{k}{M} \hat p_1\hat p_2t^2 + \frac{k^2}{3M}\hat p_2^3t^3 \\
\hat B_2 & = &\hat p_2
\end{array} \right.
\end{equation}
satisfying the Heisenberg algebra $[\hat A_1,\hat B_1]=[\hat A_2,\hat B_2]=i\hbar$, all other commutators being zero. The transformation (52) is unitary and generated by $\hat T=-\hat H$.
Applying the Weyl map $W_{(q,p)}$ to eq.(52) and comparing the result with (50) we get:
\begin{equation}
\left\{ \begin{array}{lll}
 A_1(\vec q,\vec p,t) & = & q_1(-t)\\
 B_1(\vec q,\vec p,t) & = & p_1(-t)\\
 A_2(\vec q,\vec p,t) & = & q_2(-t)\\
 B_2(\vec q,\vec p,t) & = & p_2(-t)
\end{array} \right.
\Longleftrightarrow
\left\{ \begin{array}{lll}
q_1(\vec A,\vec B,t) & = & A_1 + \frac{B_1}{M} t - \frac{k}{2M} B_2^2t^2 \\
p_1(\vec A,\vec B,t) & = & B_1-kB_2^2t \\
q_2(\vec A,\vec B,t) & = & A_2 + \left\{\frac{B_2}{m} + 2kA_1B_2 \right\}t + \frac{k}{M} B_1B_2t^2 - \frac{k^2}{3M}B_2^3t^3 \\
p_2(\vec A,\vec B,t) & = & B_2
\end{array} \right.
\end{equation}
The density matrix satisfies $\hat{\rho}(\hat{\vec q},\hat{\vec p},t)=\hat{\rho}(\hat{\vec A}(\hat{\vec q},\hat{\vec p},t),\hat{\vec B}(\hat{\vec q},\hat{\vec p},t),0)$ (cf.(36))
and thus the Wigner function $f_W(\vec A,\vec B,t)=W_{(A,B)}(\hat{\rho})=f_W(\vec A,\vec B,0)$ is static.
On the other hand, in this representation, the fundamental stargenfunctions do evolve in time. For instance (let $|x>$ be the general eigenket of $\hat q_1$ with associated eigenvalue $x$):
\begin{eqnarray}
g_{x}(\vec A,\vec B,t) & = & W_{(A,B)} (|x><x|)  =  \delta_{*_{(A,B)}} [q_1(\vec A,\vec B,t)-x]  \\
&=& \frac{1}{2\pi} \int d k \, e_{*_{(A,B)}}^{ik(q_1(\vec A,\vec B,t)-x)}=\frac{1}{2\pi} \int d k \, e^{ik(q_1(\vec A,\vec B,t)-x)}=\delta [q_1(\vec A,\vec B,t)-x] \nonumber
\end{eqnarray}
satisfies:
\begin{equation}
\frac{\partial }{\partial t} g_{x}(\vec A,\vec B,t) =  [g_{x}(\vec A,\vec B,t),H]_{M_{(A,B)}}=
\{g_{x}(\vec A,\vec B,t),H\}_{(A,B)}.
\end{equation}
Hence, the Weyl transform $W_{(A,B)}$ casts the phase space dynamics in the Heisenberg picture. Accordingly, the time dependence is exclusively displayed by the observable (stargenfunction) sector of the theory.

We now consider the action of the generalized Weyl map $W_{(q,p)}^{(A,B)}$. The associated time dependent covariant starproduct $*'_{(q,p)}$ and Moyal bracket $[\, , \, ]_{M_{(q,p)}}$ are characterized by (using the time dependent version of eqs.(6-8) and making $O'^1=p_1,\,O'^2=p_2,\, O'^3=q_1,\,O'^4=q_2$, $O^1=B_1,\,O^2=B_2,\, O^3=A_1,\,O^4=A_2$ and $i,j=1..4$):
\begin{eqnarray}
& J'^{ij}_{(q,p)} =  J^{ij}_{(q,p)} & \\
& \Gamma'^1_{22}=2kt, \quad \Gamma'^3_{22}=\frac{k}{M}t^2, \quad \Gamma'^4_{12}=\Gamma'^4_{21}= \frac{k}{M}t^2, \quad \Gamma'^4_{22}=\frac{2k^2}{M}p_2 t^3, \quad \Gamma'^4_{32}=\Gamma'^4_{23}= -2kt, & \nonumber
\end{eqnarray}
all other Christoffel symbols being zero. Notice that the connection is time dependent.

The new Wigner function (cf.(53)):
\begin{equation}
f'_W(\vec q,\vec p,t)=W_{(q,p)}^{(A,B)}(\hat{\rho})=f_W(\vec A(\vec q,\vec p,t),\vec B(\vec q,\vec p,t),0)= f_W(\vec q(\vec q,\vec p,-t),\vec p(\vec q,\vec p,-t),0)
\end{equation}
satisfies the Liouville equation:
\begin{equation}
\frac{\partial f'_W}{\partial t}=\frac{\partial f_W}{\partial \vec A} \cdot \frac{\partial \vec A}{\partial t} +
\frac{\partial f_W}{\partial \vec B} \cdot \frac{\partial \vec B}{\partial t}
=\frac{\partial f_W}{\partial \vec q} \cdot \{H,\vec q\}_{(q,p)}+\frac{\partial f_W}{\partial \vec p} \cdot \{H,\vec p\}_{(q,p)} = \{H,f_W\}_{(q,p)}=\{H,f'_W\}_{(q,p)}
\end{equation}
The quantum behavior is displayed by the stargenfunction sector alone.
However, for this system, we also have (let $z=q_1,p_1 \vee p_2$ and $|z_0>$ be a generic eigenket of $\hat z$ with associated eigenvalue $z_0$):
\begin{eqnarray}
W_{(q,p)}^{(A,B)}(|z_0><z_0|) & = & \delta_{*'(q,p)}(z-z_0)=\delta_{*(A,B)}(z(\vec A,\vec B,t)-z_0)|_{\vec A=\vec A(\vec q,\vec p, t) \wedge \vec B=\vec B(\vec q,\vec p,t)} \nonumber \\
& = & \delta(z(\vec A,\vec B,t)-z_0)|_{\vec A=\vec A(\vec q,\vec p, t) \wedge \vec B=\vec B(\vec q,\vec p,t)}= \delta(z-z_0)
\end{eqnarray}
where in the third step we used the fact that $e_{*_{A,B)}}^{ik(z(\vec A,\vec B,t)-z_0)}= e^{ik(z(\vec A,\vec B,t)-z_0)}$.
Hence, the former three fundamental stargenfunctions display a classical structure and satisfy $\{\delta_{*'(q,p)}(z-z_0),H\}=0$. We conclude that for this system, in this representation, the non-trivial (quantum) behavior is displayed by the stargenfunction $z=q_2$ alone.

A final remark is in order: in this example we were not required to use the most general formalism of section 5-(3) (Causal picture and the map $W_{(Q,P)}^{(A,B)}$) to derive the causal phase space representation of the system. This is so because the dynamical structure of the system is exceptionally simple: the quantum and the classical trajectories (which in the most general case have to be described by two different sets of coordinates, $(\vec q,\vec p)$ and $(\vec Q,\vec P)$, respectively) are identical. Indeed, eq.(50) solves both the Moyal and the Hamiltonian equations of motion and thus we were not required to introduce a second set of "classical" coordinates $(\vec Q,\vec P)$.

\section{Conclusions}

Using a time dependent extension of the generalized Weyl map we enlarged the set of possible phase space representations of quantum mechanics, derived a more general formula for the $\hbar$-deformations of the Poisson bracket and proved that there is a phase space representation where the quantum quasi-distribution displays a causal dynamical structure.
In this formulation the quantum behavior is displayed by the $*$-genfunctions (observables) sector alone. Such a property may lead to interesting applications in the field of the semiclassical limit of quantum mechanics given the fact that in the causal representation the quantum behavior has become, in fact, independent from the state of the system.

The comparison with the De Broglie Bohm formulation seems inevitable.
In the De Broglie Bohm theory the source of quantum behavior is the quantum potential together with a modification of the momentum $*$-genvalue equation. The theory admits an interpretation in terms of causal (but not classical) trajectories. In the Wigner causal formulation the effect of the quantum potential has been replaced by further corrections in the observables ($*$-genfunctions) sector of the theory and the achievement was that the particle trajectories became fully classical. It is quite remarkable that this shift of the source of the quantum behavior (from the distributional to the observables sector) could be fully performed and it may lead to an alternative causal interpretation for quantum mechanics.

\paragraph*{Acknowledgments.}

This work was partially supported by the grants POCTI/MAT/45306/2002 and POCTI/FNU/49543/2002.

\end{document}